\documentclass[aps,pre,twocolumn,superscriptaddress,showpacs]{revtex4-1}
\usepackage{graphicx}
\usepackage{amsmath}
\usepackage{natbib}
\usepackage[usenames]{color}
\usepackage{soul}

\usepackage{amsfonts}
\usepackage{dcolumn}        
\usepackage{amssymb}
\usepackage{bm,amssymb}

\usepackage{bm,fancybox}                	     

\begin{document}

\newcommand{\bea}{\begin{eqnarray}}
\newcommand{\eea}{  \end{eqnarray}}
\newcommand{\bit}{\begin{itemize}}
\newcommand{\eit}{  \end{itemize}}

\newcommand{\be}{\begin{equation}}
\newcommand{\ee}{\end{equation}}
\newcommand{\ra}{\rangle}
\newcommand{\la}{\langle}
\newcommand{\U}{\widetilde{U}}

\def\bra#1{{\langle#1|}}
\def\ket#1{{|#1\rangle}}
\def\bracket#1#2{{\langle#1|#2\rangle}}
\def\inner#1#2{{\langle#1|#2\rangle}}
\def\expect#1{{\langle#1\rangle}}
\def\e{{\rm e}}
\def\proj{{\hat{\cal P}}}
\def\tr{{\rm Tr}}
\def\H{{\hat H}}
\def\Hdag{{\hat H}^\dagger}
\def\Lop{{\cal L}}
\def\Ehat{{\hat E}}
\def\Edag{{\hat E}^\dagger}
\def\Shat{\hat{S}}
\def\Sdag{{\hat S}^\dagger}
\def\Ahat{{\hat A}}
\def\Adag{{\hat A}^\dagger}
\def\U{{\hat U}}
\def\Udag{{\hat U}^\dagger}
\def\Zhat{{\hat Z}}
\def\Phat{{\hat P}}
\def\Op{{\hat O}}
\def\id{{\hat I}}
\def\x{{\hat x}}
\def\P{{\hat P}}
\def\Px{\proj_x}
\def\Pr{\proj_{R}}
\def\Pl{\proj_{L}}


\title{Short periodic orbits theory for partially open quantum maps}

\author{Gabriel G. Carlo}
\email[E--mail address: ]{carlo@tandar.cnea.gov.ar}
\affiliation{Comisi\'on Nacional de Energ\'{\i}a At\'omica, CONICET, 
Departamento de F\'{\i}sica. 
Av.~del Libertador 8250, 1429 Buenos Aires, Argentina}

 \author{R. M. Benito}
\affiliation{Grupo de Sistemas Complejos
 and Departamento de F\'{\i}sica,
 Escuela T\'ecnica Superior de Ingenieros
 Agr\'onomos, Universidad Polit\'ecnica de Madrid,
 28040 Madrid, Spain}

\author{F. Borondo}
\affiliation{Departamento de Qu\'{\i}mica, and
 Instituto de Ciencias Matem\'aticas (ICMAT),
 Universidad Aut\'onoma de Madrid,
 Cantoblanco, 28049--Madrid, Spain}
\date{\today}
\begin{abstract}

We extend the semiclassical theory of short periodic orbits 
[Phys. Rev. E {\bf 80}, 035202(R) (2009)] to partially 
open quantum maps. They correspond to classical maps where the trajectories are partially 
bounced back due to a finite reflectivity $R$. These maps are representative of a 
class that has many experimental applications. The open scar functions are conveniently redefined, providing 
a suitable tool for the investigation of these kind of systems. Our theory is applied 
to the paradigmatic partially open tribaker map. 
We find that the set of periodic orbits that belong to the classical repeller 
of the open map ($R=0$) are able to support the set of long-lived resonances of the partially open quantum 
map in a perturbative regime. By including the most relevant trajectories outside of this 
set, the validity of the approximation is extended to a broad range of $R$ values. 
Finally, we identify the details of the transition from qualitatively 
open to qualitatively closed behaviour, providing an explanation in terms of short periodic orbits.
\end{abstract}
\pacs{05.45.Mt, 03.65.Sq}
\maketitle

\section{Introduction}
 \label{sec:intro}

In completely open quantum systems the fractal Weyl law \cite{conjecture} states that the number of long-lived resonances 
scales with the Planck constant as $\hbar^{-d/2}$, where $d+1$ is the fractal dimension
of the classical repeller. This has been thoroughly tested \cite{Nonnenmacher} and it is 
a well established result \cite{Hamiltonians,qmaps}. At the center-stage we find the classical 
invariant distribution consisting of all the trajectories that do not escape either in the past 
or in the future: the repeller. It plays a fundamental role in the determination of the quantum spectral features. 
However, a finite reflectivity $R$ is required to describe many experimental situations, as it is the case 
in optical cavities \cite{WiersigR,Microlasers} for 
example. This means that the 
classical trajectories arriving to the opening are partially reflected. 
This leads to partially open systems which can be very well represented with 
partially open maps. In these cases we have that the fractal dimension is the phase space dimension, but 
the fractal character remains through the multifractal measures \cite{Ott}. 
In a very recent work \cite{Altmann} the case of the partially open tribaker map has been analyzed. 
It was found that the number of long-lived resonances follows a non trivial scaling that is related to 
the quantum undersampling of the classical phase space.

On the other hand, the semiclassical theory of short periodic orbits (POs) has been successfully applied to 
a number of problems including closed quantum chaotic systems, scarring phenomena and more recently to 
open quantum maps \cite{art1}. In this approach the main ingredients are the shortest POs contained in the 
repeller, they provide all the necessary information to construct a basis set of scar functions in which the 
quantum non unitary operators can be written. The number of trajectories needed to reproduce the quantum 
repeller \cite{art0,art3} is related to the fractal Weyl law \cite{art2}. 

In this work we extend the short POs theory to partially open quantum maps where 
a fraction of the quantum probability is reflected. We apply it to the partially open tribaker map. 
For this purpose we modify the definition of the scar functions 
and define the way in which the trajectories inside and outside of the repeller at $R=0$ are taken into 
account. We find that there are several regimes as a function of the reflectivity. 
First of all, a perturbative one in which by just considering the 
shortest POs inside the repeller at $R=0$ the long-lived resonances and the quantum distribution associated to the 
invariant classical measure can be obtained.  
By suitably incorporating the shortest POs outside 
of this set we are able to extend the validity of the semiclassical calculations well beyond this 
perturbative regime. In doing so we keep some of its efficiency in terms of reducing the dimension of 
the Hilbert space needed for the calculations. Finally we find that a transition from an open-like to a closed-like 
behavior takes place and this is clearly characterized by its effect on the 
semiclassical calculations.
 
This paper adopts the following organization: 
In Sec. \ref{sec:POpenmaps} we define the classical and quantum partially open tribaker maps and all the relevant 
quantities associated to them. 
In Sec. \ref{sec:shortPOth2POmaps} we extend the short PO theory to this kind of systems.
In Sec. \ref{sec:results} we apply it to our map and discuss the results.
We conclude in Sec. \ref{sec:conclusions}

\section{Partially open maps and the tribaker example}
 \label{sec:POpenmaps}

The study of maps, which constitute simple examples exhibiting a rich dynamics, has a long and fruitful history 
in the classical and quantum chaos literature \cite{Ozorio 1994,Hannay 1980,Espositi 2005}. 
In this respect, open maps on the 2-torus are transformations 
which represent the evolution of trajectories that disappear when they 
reach an open region in the bidimensional phase-space. An invariant set is formed by the 
remaining trajectories, i.e. those that do not escape either in the past or in the future. 
These trajectories build the forward and backwards trapped sets respectively, and the 
intersection of both is what is called the repeller, which has a fractal dimension. 

Partially open maps can be defined as those maps in which the opening does not absorb 
all the trajectories that arrive at it, but reflects back a certain amount. 
This amount is essentially given by the reflectivity $R \in (0:1)$. Here, we exclude 
$R=0$ since this corresponds to completely open maps, and $R=1$ since it represents 
a closed one. This is the simplest choice of the reflection 
mechanism. In general one considers a function of the phase space points, $R(q,p)$ in 
bidimensional examples. Nevertheless our simple model captures the main features of realistic 
systems of interest such as microcavities \cite{Altmann}.

In contrast to what happens in an open map, in partially open ones the relevant measure 
is not uniformly distributed on a fractal, showing multifractality instead.
We closely follow the definition found in \cite{Altmann}. 
In each phase space region $X_i$ this measure depends on 
the average intensity $I_t$ with $t \rightarrow \infty$ of random initial conditions taken in 
$X_i$. These intensities are defined as $I_{t+1}=R I_t$ with $I_0=1$ for each trajectory. 
The finite time measure of $X_i$ can be defined as $\mu_{t,i}^{b}=\langle I_{t,i} \rangle/\sum_i 
\langle I_{t,i} \rangle$ where the average is over the initial conditions in the given phase space region. 
In fact, this measure is the analogue of the backwards trapped set of open maps. 
If we evolve backwards we obtain $\mu_{t,i}^{f}$ the analogue of the forward trapped set, and the intersection 
gives what we call the {\em partial repeller} $\mu_{t,i}$.
 
The usual quantization scheme for maps on the torus proceeds in the following way: 
in the first place we impose boundary conditions for
both the position and momentum representations by taking 
$\bracket{q+1}{\psi}\:=\:e^{i 2 \pi \chi_q}\bracket{q}{\psi}$, and
$\bracket{p+1}{\psi}\:=\:e^{i 2 \pi \chi_p}\bracket{p}{\psi}$, with $\chi_q$, $\chi_p \in
[0,1)$. Thus the Hilbert space is of finite dimension $N=(2 \pi \hbar)^{-1}$, and the
semiclassical limit corresponds to $N \rightarrow \infty$. The system's propagator is given by a
$N\times N$ matrix. Position and momentum eigenstates are given by
$\ket{q_j}\:=\:\ket{(j+\chi_q)/N}$ and $\ket{p_j}\:=\:\ket{(j+\chi_p)/N}$ with
$j\in\{0,\ldots, N-1\}$. A discrete Fourier transform gives 
$\bracket{p_k}{q_j}\:=\: \frac{1}{\sqrt{N}} e^{-2i\pi(j+\chi_q)(k+\chi_p)/N} \: \equiv \:
(G^{\chi_q, \chi_p}_N)$.

When $R=0$ the opening can be easily quantized as a projection operator $P$ on its
complement. We usually take a finite strip parallel to the $p$ axis, so if $U$ is the propagator 
for the closed system, then $\widetilde{U}=PUP$ stands for the open one. 
Here we take a partial opening so we modify this projector by replacing 
the zero block of the opening by $\sqrt{R} \times \openone$, where the identity has the dimension 
associated to the escape region. 
The resulting partially open quantum map has $N$ right eigenvectors $|\Psi^R_j\ra$ and $N$ left
ones $\la \Psi_j^L|$, which are mutually orthogonal $\la
\Psi_j^L|\Psi^R_k\ra=\delta_{jk}$, and that are associated to resonances $z_j$. We choose $\la
\Psi_j^R|\Psi^R_j\ra=\la \Psi_j^L|\Psi^L_j\ra$ for the norm.

We make all the calculations of this work on the tribaker map, whose 
classical expression is given by 
\begin{equation}
\mathcal B(q,p)=\left\{
  \begin{array}{lc}
  (3q,p/3) & \mbox{if } 0\leq q<1/3 \\
  (3q-1,(p+1)/3) & \mbox{if } 1/3\leq q<2/3\\
  (3q-2,(p+2)/3) & \mbox{if } 2/3\leq q<1\\
  \end{array}\right.
\label{classicaltribaker}
\end{equation}
This is an area-preserving, uniformly hyperbolic, piecewise-linear and invertible map with
Lyapunov exponent $\lambda=\ln{3}$. An opening has been placed in the region $1/3< q<2/3$, where the 
reflectivity is given by $R$ as explained above.

The quantum version of the tribaker map is defined by means of the 
discrete Fourier transform in position representation as \cite{Saraceno1,Saraceno2}
\begin{equation}\label{quantumbaker}
 U^{\mathcal{B}}=G_{N}^{-1} \left(\begin{array}{ccc}
  G_{N/3} & 0 & 0\\
  0 & G_{N/3} & 0\\
  0 & 0 & G_{N/3}\\
  \end{array} \right),
\end{equation}
taking antiperiodic boundary conditions, this meaning $\chi_q=\chi_p=1/2$. The partially open quantum tribaker map 
is then given by means of the operator 
\begin{equation}\label{partialprojector}
 P=\left(\begin{array}{ccc}
  \openone_{N/3} & 0 & 0\\
  0 & \sqrt{R} \openone_{N/3} & 0\\
  0 & 0 & \openone_{N/3}\\
  \end{array} \right),
\end{equation}
 applied to Eq. (\ref{quantumbaker}), thus obtaining 
\begin{equation}\label{partiallyopenquantumbaker}
 \widetilde{U^{\mathcal{B}}}= P U^{\mathcal{B}} P.
\end{equation}

\begin{figure}
\includegraphics[width=8cm]{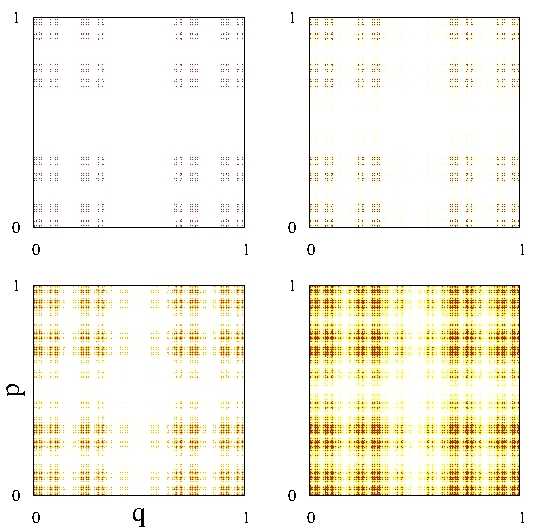}
\caption{(color online) Classical measure $\mu_{t,i}$, i.e. the partial repeller 
 on the 2-torus for the partially open tribaker map, for four different values of the reflectivity, R. 
 In the upper left panel we show the $R=0$ case. In the upper right panel 
 we can observe the $R=0.01$ case which shows no appreciable differences with respect 
 to the previous value of the reflectivity. The lower left panel corresponds to $R=0.07$, 
 where we can find a finite measure outside of the repeller and finally the 
 lower right panel for $R=0.2$ where the measure now extends to almost all regions of phase space.}
\label{fig1}
\end{figure}

In Fig \ref{fig1} we show the finite time partial repeller $\mu_{t,i}$ at time $t=10$. We have 
selected four representative examples for the reflectivity. In the upper left panel we take $R=0$ for comparison purposes. 
Next, in the upper right panel we display the $R=0.01$ case which is almost indistinguishable 
from the completely open one, this suggests that there should be a perturbative 
regime at least up to these reflectivity values. In the lower left panel we represent the case 
$R=0.07$, where it is clear that the measure starts to be non negligible outside the repeller 
($R=0$). Finally in the lower right panel the $R=0.2$ case underlines the need to consider a 
much different scenario with widespread finite measure over the 2-torus.

\section{Short Periodic Orbits theory for partially open quantum maps}
 \label{sec:shortPOth2POmaps}

Following the ideas of the short POs theory for closed systems \cite{Vergini}, 
we have recently developed a similar theory of short POs for open quantum maps \cite{art1,art2}. 
In this theory the repeller has a central role and the short POs that belong to it provide all the 
essential information needed to recover the quantum long-lived eigenvalues and the 
quantum repeller (with some exceptions \cite{art3}). The fundamental tools in this 
approach are the open scar functions associated to each one of these trajectories. 
To make the paper self-contained, we make a brief description of the partially open scar function construction, 
which is a very natural adaptation to the case of partially open maps.

Let $\gamma$ be a PO of fundamental period $L$ that belongs to a partially open map. 
We can define coherent states $|q_j,p_j\rangle$ associated to each point of the orbit 
(it has a total of $L$ points, all in the partial repeller). We then construct a linear combination 
with them: \begin{equation} |\phi_\gamma^m\rangle=\frac{1}{\sqrt{L}}\sum_{j=0}^{L-1}
\exp\{-2\pi i(j A^m_\gamma-N\theta_j)\}|q_j,p_j\rangle,\end{equation} where 
$m\in\{0,\ldots,L-1\}$ and $\theta_j=\sum_{l=0}^j S_l$. In this expression $S_l$ is the action acquired 
by the $l$th coherent state in one step of the map. The total action is $\theta_L\equiv S_\gamma$
and $A^m_\gamma=(NS_\gamma+m)/L$. Finally, the right and left scar functions for the periodic 
orbit are defined through the propagation of these linear combinations under the partially open map 
$\widetilde{U}$ (up to approximately the system's Ehrenfest time $\tau$). 
\begin{equation}\label{prop}
|\psi^R_{\gamma,m}\rangle=\frac{1}{\mathcal{N}_\gamma^R}\sum_{t=0}^{\tau}
\widetilde{U}^te^{-2\pi iA^m_\gamma t}\cos\left(\frac{\pi
t}{2\tau}\right)|\phi_\gamma^m\rangle,\end{equation} and \begin{equation}
\langle\psi^L_{\gamma,m}|=\frac{1}{\mathcal{N}_\gamma^L}\sum_{t=0}^{\tau}
\langle\phi_\gamma^m|\widetilde{U}^te^{-2\pi iA^m_\gamma t}\cos\left(\frac{\pi
t}{2\tau}\right).\end{equation} Normalization ($\mathcal{N}_{\gamma}^{R,L}$) 
is chosen in such a way that $\langle \psi_{\gamma,m}^R|\psi^R_{\gamma,m}\rangle=
\langle \psi_{\gamma,m}^L|\psi^L_{\gamma,m}\rangle$ and
$\langle \psi_{\gamma,m}^L|\psi^R_{\gamma,m}\rangle=1$.
These functions are suitable tools for the investigation of the morphology 
of the eigenstates.

\begin{figure}
\includegraphics[width=8cm]{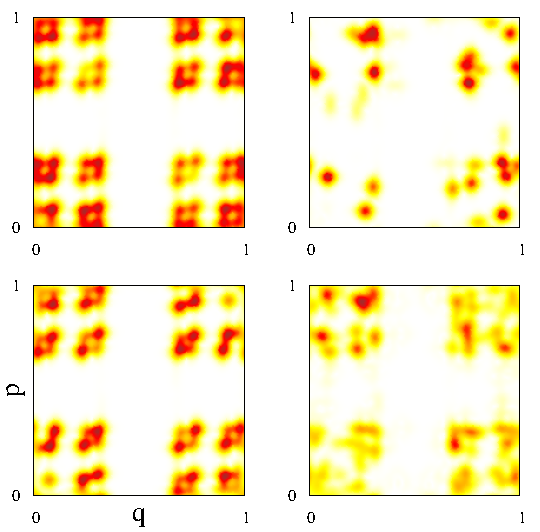}
\caption{(color online) Sum of $\hat{h}_j$ over the partially open scar functions. Upper panels correspond 
to $R=0.07$ and lower ones to $R=0.2$. On the left column we show the scar functions set associated 
to orbits $\gamma$ inside the repeller at $R=0$, on the right one the set associated to orbits 
outside of it.}
\label{fig2}
\end{figure}

In Fig. \ref{fig2} we illustrate the partially open scar functions by means of 
a representation \cite{art0} that clearly shows the quantum probability 
that can be associated to the classical partial repeller. 
We define the symmetrical operator $\hat{h}_j$ related to the right 
$\vert \psi^R_j\rangle$ and left $\langle \psi^L_j\vert$ states (in this case scar functions, where 
we have collapsed both subscripts to just one for simplicity)
\begin{equation}\label{eq.hdef}
 \hat{h}_j=\frac{\vert \psi^R_j\rangle\langle \psi^L_j\vert}{\langle \psi^L_j\vert \psi^R_j\rangle},
\end{equation}
which is associated to the orbit $\gamma$. By calculating the sum over all these projectors \cite{art3} 
corresponding to the sets of scar functions used for a given semiclassical calculation we can 
see how different parts of the phase space are represented in the basis. In the upper panels of Fig. \ref{fig2} 
we show these sums for  $R=0.07$, while in the lower ones for $R=0.2$. On the left column we find open scar 
functions associated to short POs living inside the repeller, and outside of 
it on the right column. These sets were actually used for some of the calculations to be described in Sec. \ref{sec:results}. 
It can be noticed that the contribution from the orbits that are outside of the repeller increases with $R$.

As a matter of fact, we come now to the other important point in adapting the theory to the partially 
open case: how to select the orbits 
that take part in the calculation? For this purpose we choose the following criterion: 
select a given number of POs, $N^{POs}$, from the whole set up to a period $L$, 
that approximately covers the partial repeller. The selected degree in the approximation of this 
covering translates into the fraction of long-lived resonances that can be obtained.
In practical terms this means allowing all POs up to period $L$ that are 
inside the repeller into a first list, since they have a uniform weight. 
Those that are outside must have the greatest values of $\mu$.  
We select them by fixing the maximum number allowed $N_{max}^{outPO}$ (again, from all of them up to period $L$), 
which establishes a $\mu$ cutoff value from the list of these orbits ordered by decreasing weight. 
$N_{max}^{outPO}$ is established having in mind that it should increase with $R$, 
starting from zero at $R=0$. For that purpose $N_{max}^{outPO}$ grows with the reflectivity 
(note that this also depends on the particular map considered). 
Hence, the preliminary list of all the orbits incorporates them by increasing value of their period 
and contains the ones with the largest weights. We notice that the number of orbits in this list is typically 
much larger than the number needed for our approximation. As a next step, we optimize the set by a reordering 
to provide with the most uniform covering possible of the partial repeller. 
We finish the selection by cutting the list at $N^{POs}$.  

Finally, we construct an appropriate basis in which we can write the partially open evolution operators 
associated to partially open maps as 
$\langle\psi_{\alpha,i}^L|\widetilde{U}|\psi_{\beta,j}^R\rangle$. This expression is the short POs approximation to the 
partially open propagator $\widetilde{U}$ on the partial repeller. 
Equipped with $\langle \psi_n^L|\psi_m^R\rangle\neq\delta_{nm}$, 
we solve a generalized eigenvalue problem that provides the eigenstates of this matrix. 
The long-lived resonances \cite{art1} are constructed by a linear 
combination of the eigenvector's coefficients and the corresponding scar functions of the basis. 

\section{Results}
 \label{sec:results}

We use the short POs theory to construct a semiclassical approximation of the 
partially open quantum tribaker map in the $N=243$ case, for several 
values of $R$, considering POs up to period $L=7$. In order to quantify the behaviour of the short POs method 
we define its performance $P$ \cite{art3} as the fraction of long-lived 
eigenvalues that it is able to reproduce within an error given by
$\epsilon=\sqrt{({\rm Re}{(z_i^{ex})}-{\rm Re}{(z_i^{sc})})^2+
({\rm Im}{(z_i^{ex})}-{\rm Im}{(z_i^{sc})})^2}$, 
where $z_i^{ex}$ and $z_i^{sc}$ are the exact eigenvalues and those given 
by the semiclassical theory, respectively. 
We restrict our analysis to the number of exact eigenvalues with modulus greater than $\nu_c$, which 
is a critical value that depends on $R$. In our calculations we have tried to keep the number 
constant at around $n_c=60$ since for low values of $R$ this represents the outer ring of 
eigenvalues that is a typical feature of the open quantum baker maps. This is very useful 
since it provides with a natural separation between short and long-lived resonances. 
We calculate the number of scar functions $N_{SF}$ as a fraction of $N$ that are 
needed in order to obtain as many semiclassical eigenvalues inside the $\epsilon=0.001$ vicinity of the 
corresponding exact ones in order to reach $P \geq 0.8$. 
The fraction $N_{SF}/N$ is a good indicator of the efficiency of the method. In fact, one of its main advantages 
is reducing the effective dimension of the matrices that one needs to diagonalize in order 
to obtain the resonances of the quantum system. But it also reveals if there are quantum signatures 
of (multi)fractal dimensions when partially opening the map (like the fractal Weyl law). 
It is worth mentioning that all the threshold values considered guarantee a reasonably 
good performance of the approximation and the evaluation of a meaningful number of eigenvalues 
in the whole range of situations that we have studied. 

\begin{figure}
  \includegraphics[width=8cm]{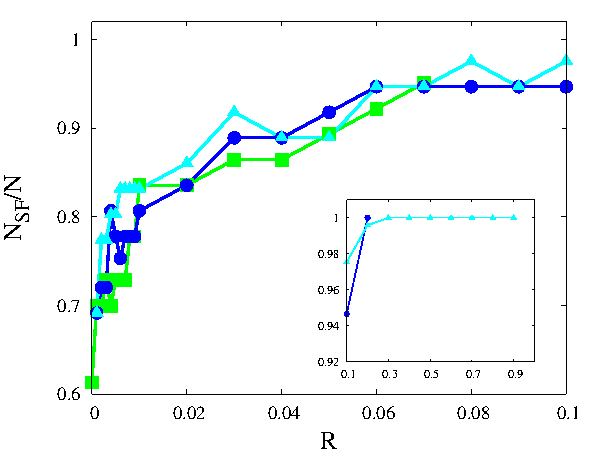}
 \caption{(color online)  Fraction of scar functions $N_{SF}/N$ needed to reach 
 $P=0.8$ as a function of the reflectivity $R$. The green (gray) line with squares corresponds 
 to the case where only POs inside the repeller have been considered. Blue (black) and 
 Cyan (light gray) lines with circles and up triangles correspond to considering $N_{max}^{outPO}=5$ and 
 $N_{max}^{outPO}=50$, respectively. The inset shows the behaviour for greater $R$ values.
}
 \label{fig3}
\end{figure}

In Fig. \ref{fig3} we show the fraction $N_{SF}/N$ needed to reach $P \geq 0.8$ 
as a function of $R \in [0:0.1]$ for three different scenarios derived from our POs selection 
criterion. The line with squares correspond 
to the case in which we only take POs that belong to the repeller. We have also considered 
$N_{max}^{outPO}=5$ (POs outside of it), results which are represented by means of a line 
with circles. Finally the case with $N_{max}^{outPO}=50$ is shown through a line with up triangles. 
We point out that there is no improvement in the calculations when considering more POs outside the repeller.

The results in Fig. \ref{fig3} clearly show that the behaviour of the short POs theory for the partially open tribaker map 
can be divided into four regimes. First we notice 
that there is a perturbative regime in which the open scar functions associated to the 
POs inside the repeller at $R=0$ are the only ones needed in order to accurately reproduce the 
long-lived resonances. 
This regime extends up to approximately $R=0.01$. It is also clear that by incorporating up to 
$5$ POs outside the repeller with the greatest 
$\mu$ values the performance is worse. 
This amount represents around $1\%$ of  
the available POs up to period $L=7$. Moreover, if we include up to $50$ of these POs 
the result is much worse (a significant amount of the POs inside the repeller are now replaced by them). 
Beyond $R=0.01$ the POs in the repeller are still enough to reproduce the long-lived resonances, but 
the overall amount of scar functions needed steadily increases up to $R=0.07$, though at a slower pace 
than before. In fact, for $R \in [0.01:0.07]$ 
we identify a semi-perturbative regime, in the sense that the classical information of the repeller is still 
enough but becomes less efficient. Above $R=0.07$ the repeller is not enough to adequately treat the problem 
and other POs outside of it are needed. From this value of $R$ on, we also see that by incorporating 
a greater amount of POs not living inside the repeller, 
we obtain a similar performance of the method.  
This so to say proper partially open regime extends up to $R=0.3$ approximately. 
Beyond that we find the fourth and last regime, as can be seen in the inset of Fig. \ref{fig3}. 
Here, a saturation occurs and we need $N_{SF}=N$, leaving our method with no advantage with respect to a direct diagonalization. 
This is a clear sign that the system has become effectively closed from our perspective, 
i.e. we need to expand all the phase space no matter the local value of the measure $\mu$. 
We have verified that in the case $N=729$, $N_{SF}/N=0.69$ for $R=0.01$ obtained with all 
the POs inside the repeller, and $N_{SF}/N=0.96$ for $R=0.1$ with POs mainly outside of it. 
This is in agreement with the previous results, though we leave the study of this scaling for future work.

To illustrate this behaviour we use the projectors $\hat{h}_j$ of Sec. \ref{sec:shortPOth2POmaps}
now associated to the right $|\Psi^R_j\ra$ and left $\la \Psi^L_j|$ eigenstates, 
which are related to the eigenvalue $z_j$. We calculate the sum of the 
first $j$ of these projectors \cite{art3}, ordered by decreasing modulus of the corresponding eigenvalues  
($\vert z_j\vert\geqslant\vert z_{j^\prime}\vert$ with $j\leq j^\prime$) up to completing the set of 
long-lived resonances. 
\begin{equation}
\hat{Q}_j\equiv\sum_{j^\prime=1}^j\hat{h}_{j^\prime}.
\end{equation}
Their phase space representation by means of 
coherent states $\vert q,p\rangle$ is given by 
\begin{eqnarray}
 h_j(q,p)&=&\vert\langle q,p\vert \hat{h}_j\vert q,p\rangle\vert\\
 Q_j(q,p)&=& \vert\langle q,p\vert \hat{Q}_j\vert q,p\rangle\vert. 
\end{eqnarray}
This is our formal definition of the {\em partial quantum repeller}, which we call $Q_{n_c}$.

\begin{figure}
  \includegraphics[width=8cm]{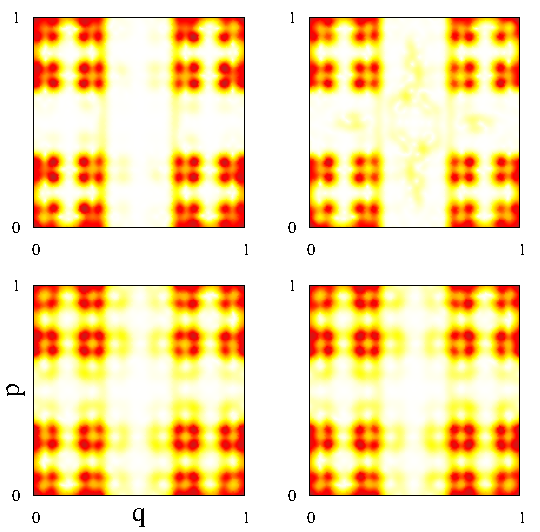}
 \caption{(color online) $Q_{n_c}$ for the exact resonances (left column), 
 and $Q_{n_c}^{sc}$ for the short POs theory results (right column), 
 for $R=0.07$ (top) and  $R=0.2$ (bottom).}
 \label{fig4}
\end{figure}

In Fig. \ref{fig4} we show $Q_{n_c}$ for the exact resonances on the left column, and 
the ones given by the short POs approach, i.e. $Q_{n_c}^{sc}$, on the right one. 
In the upper panels the case $R=0.07$ has been obtained by just using the POs that live in the 
repeller. This shows that the POs inside the repeller are 
enough to reproduce the partial quantum repeller; little probability is found 
outside of it. In the lower panel we display 
the $R=0.2$ case where the contribution from the orbits outside of the repeller is crucial, as 
can be noticed from the greater non-zero probability found in other regions of 
phase space.
The overlap of the normalized distributions, calculated as $O=\iint Q_{n_c}(q,p) Q_{n_c}^{sc}(q,p) dq dp$, 
is $O=0.994$ for $R=0.07$ and $O=1.0$ for $R=0.2$, confirming the excellent agreement between 
the exact and the semiclassical results.

\section{Conclusions}
 \label{sec:conclusions}

 We have extended the short POs theory to partially open quantum maps and have applied it to the 
 particular case of the partially open tribaker map. It turns out that for low reflectivities, the 
 long-lived resonances and the partial quantum repeller 
 can be reproduced up to a very good accuracy with just the classical 
 information of the repeller ($R=0$). Moreover, we identify 
 four different regimes in which the role played by the orbits that live inside and outside of 
 the repeller changes. They are the perturbative, semi-perturbative, proper partially open, 
 and the effectively closed regimes. This translates directly into the partial quantum repeller, 
 which performs a non-trivial transition from an open-like to a closed-like 
 shape as a function of $R$.
 
 One of the main features of our theory when applied to open maps is that 
 the size of the matrices needed in the final diagonalization is greatly reduced. This is due 
 to the fact that the number of necessary scar functions scales according with the fractal Weyl law \cite{art2}. 
 In the case of the partially open tribaker map we have found that for $R<0.01$ (perturbative 
 regime) the reduction in the calculation effort is very important. 
 This advantage is gradually lost as the reflectivity $R$ increases. 
 This is in agreement with the behaviour already found in previous publications \cite{Altmann} 
 where the multifractal nature of these classical systems has been shown to influence 
 the spectral properties of their quantum counterparts. This reveals the importance 
 that all the probability distributed in each region of phase space has in this kind of maps 
 (despite we have selected the orbits in terms of their weights expressed by $\mu$). 
 Moreover, from the semiclassical point of view all the POs are progressively more interconnected as 
 a function of $R$, and each detail of them becomes more and more relevant for a precise diagonalization. 
 
 Finally, the partially open scar function is a very interesting tool 
 for the study of the morphology of the eigenstates. In fact, we think that our semiclassical 
 theory can significantly contribute and provides a new perspective for the study of emission problems 
 in microcavities.
 
\section{Acknowledgments}
The research leading to these results has received funding from CONICET (Argentina) under 
project PIP 112 201101 00703, and the  Ministerio de Econom\'\i a y Competitividad (MINECO) 
under contracts MTM2012-39101 and MTM2015-63914-P, and by ICMAT Severo Ochoa SEV-2015-0554.

%

\begin{thebibliography}
\eprint{}

\bibitem{conjecture} 
W.T. Lu, S. Sridhar and M. Zworski,
Phys. Rev. Lett. {\bf 91}, 154101 (2003).

\bibitem{Nonnenmacher} 
S. Nonnenmacher, 
arXiv:1105.2457 (2011).

\bibitem{Hamiltonians} 
J.A. Ramilowski, S.D. Prado, F. Borondo and D. Farrelly, 
Phys. Rev. E {\bf 80}, 055201(R) (2009);
A. Ebersp\"acher, J. Main and G. Wunner, 
Phys. Rev. E {\bf 82}, 046201 (2010).

\bibitem{qmaps} 
H. Schomerus and J. Tworzydlo, 
Phys. Rev. Lett. {\bf 93}, 154102 (2004);
S. Nonnenmacher and M. Rubin, 
Nonlinearity {\bf 20}, 1387 (2007); 
D. L. Shepelyansky,
Phys. Rev. E {\bf 77}, 015202(R) (2008).

\bibitem{WiersigR}
J. Kullig and J. Wiersig, 
New J. Phys. {\bf 18}, 015005 (2016).

\bibitem{Microlasers}
W. Fang, Phys. Rev. A {\bf 72}, 023815 (2005); J.U. N\"ockel and
D.A. Stone, Nature (London) {\bf 385}, 45 (1997); T. Harayama,
P. Davis and K.S. Ikeda, Phys. Rev. Lett. {\bf 90}, 063901 (2003);
J. Wiersig and M. Hentschel, Phys. Rev. A {\bf 73}, 031802(R) (2006);
J. Wiersig and M. Hentschel, Phys. Rev. Lett. {\bf 100}, 033901 (2008).

\bibitem{Ott}
E. Ott, {\it Chaos in Dynamical Systems} (Cambridge University Press, 
Cambridge, 2002), 2nd ed.

\bibitem{Altmann}
M. Sch\"onwetter and E.G. Altmann, 
Phys. Rev. E {\bf 91}, 012919 (2015).

\bibitem{art1} 
M. Novaes, J.M. Pedrosa, D. Wisniacki, G.G. Carlo, and J.P. Keating, 
Phys. Rev. E {\bf 80}, 035202(R) 2009.

\bibitem{art0} 
L. Ermann, G.G. Carlo, and M. Saraceno, 
Phys. Rev. Lett. {\bf 103}, 054102 (2009).

\bibitem{art3} 
G.G. Carlo, D.A. Wisniacki, L. Ermann, R.M. Benito, and F. Borondo, 
Phys. Rev. E {\bf 87}, 012909 (2013).

\bibitem{art2} 
J.M. Pedrosa, D. Wisniacki, G.G. Carlo, and M. Novaes, 
Phys. Rev. E {\bf 85}, 036203 (2012).

\bibitem{Ozorio 1994}
M. Basilio De Matos, A. M. Ozorio De Almeida, Ann. Phys. {\bf 237}, 46-65 (1995).

\bibitem{Hannay 1980}
J. H. Hannay, M. V. Berry, Physica D 1 267 (1980).

\bibitem{Espositi 2005}
M. Degli Espositi, B. Winn, J.Phys.A: Math.Gen.{\bf 38}, 5895-5912 (2005).

\bibitem{Saraceno1}
M. Saraceno, Ann. Phys. \textbf{199}, 37 (1990); M. Saraceno and R. O. Vallejos,
Chaos \textbf{6}, 193 (1996); A. \L ozi\'{n}ski, P. Pako\'{n}ski and K. \.{Z}yczkowski,
Phys. Rev. E \textbf{66}, 065201(R) (2002).

\bibitem{Saraceno2}
M. Saraceno and A. Voros, Physica D \textbf{79}, 206 (1994).

\bibitem{Vergini} E. G. Vergini  J. Phys. A: Math. Gen. {\bf 33} 4709 (2000);
E. G. Vergini and G. G. Carlo, J. Phys. A: Math. Gen. {\bf 33} 4717 (2000);
E. G. Vergini, D. Schneider and A. F. Rivas, J. Phys. A: Math. Theor. {\bf 41} 405102 (2008); 
L. Ermann and M. Saraceno, Phys. Rev. E {\bf 78}, 036221 (2008).

%
\end{thebibliography}
\end{document}